\documentstyle[epsfig,twocolumn,aps]{revtex}
\newcommand {\BM}[1] {\mbox{\boldmath$#1$}}
\newcommand {\BMnabla} {\BM{\nabla}}
\newcommand {\BMalpha} {\BM{\alpha}}

\newcommand {\BMomega} {\BM{\omega}}
\newcommand {\BMeta} {\BM{\eta}}
\begin{document}

\author{R.J.S. Lee, J.V. Mullan, J.F. McCann and D.S.F. Crothers}
\address{Department of Applied Mathematics $\&$ Theoretical Physics, The Queen's University of Belfast,
 \\Belfast, BT7 1NN, Northern Ireland, UK.}
\date{\today}
\title{Pair production and electron capture in relativistic heavy-ion
collisions}
\maketitle
\begin{abstract}
Results are presented for simulations of
 electron-positron pair production in relativistic heavy-ion collisions
leading to electron capture and positron ejection. We apply a two-center relativistic continuum 
distorted-wave model to represent the electron/positron dynamics during the collision process. 
The  results are compared with experimental cross-section data for 
${\rm La}^{57+}$ and ${\rm Au}^{79+}$ impact on Gold, Silver and Copper targets. The theory
is in very good agreement with experiment, verifying the result that the process increases in importance with both 
collision energy and target atomic number, and improves upon  previous simulations of this process.
\end{abstract}

Early theoretical work on the production of an $e^{-}/e^{+}$ pair through
heavy-ion collisions considered only the creation of the electron
and positron in the continuum. However {\it capture by pair production} (CPP), in which
the electron is formed in a bound state of one or other ion, becomes a significant
process at highly relativistic energies.
Remarkably this process was sufficiently important to enable the synthesis of  atomic 
antihydrogen using the low-energy antiproton ring at CERN. A beam of fast antiprotons impacting on a 
xenon gas target \cite{baur96} led to pair production with positron capture.
Theory had predicted \cite{bert88,baltz91} 
that cross sections for CPP would increase with energy, and indeed this 
has been verified experimentally  \cite{belk93,belk94a,krau98} . In fact, this 
process  eventually becomes the dominant mechanism for charge exchange
in highly-relativistic atomic collisions \cite{krau98}.   
As well as being an interesting area of study in its own right, this process has
important applications in the physics of heavy ion colliders such as the
large-hadron collider (LHC) and the relativistic heavy-ion collider (RHIC) \cite{belk98}.
The process of CPP will lead to  depletion of the charge state of the beam and hence
a loss in luminosity of the collider. For typical operating conditions of such
facilities, these losses might amount to 50\% \cite{belk98} or more.

Although the process is strongly coupled at high energy, 
simulations based on relativistic coupled-channel calculations
 \cite{baltz93,baltz94} 
have indicated that leading-order perturbation theory is adequate for total cross section estimates
for energies ($E$) up to $150\ {\rm GeV/u}$ \cite{krau98}.
 Nonetheless in the energy range  $ E \sim 1\ {\rm GeV/u}$ 
where reliable experimental data exist, theory and experiment have been in least agreement. It is
this region which we address in this paper.

It is now some thirteen years since Becker and co-workers \cite{beck87a,beck87b,anho87} 
obtained the first estimates of cross sections for pair creation 
with simultaneous capture of the electron into
the $K$-shell of one of the colliding ions. 
However with the exception of Deco and
Rivarola, who gave a two-center description of the continuum positron \cite{deco88b},
two somewhat artificial modes of reaction have been distinguished and treated
separately when modeling this process: excitation from the negative
energy continuum of an ion to one of its bound states 
 \cite{bert88,beck87b,deco88a}
or transfer to a bound state of the other ion 
\cite{eich95,ione96}. Such approaches, while suited to  circumstances in
which one ion is much more
highly charged than the other, lack symmetry and make a distinction
between two separate modes of CPP. They lead to different formulae within
first-order perturbation theory \cite{eich95} and hence different projectile charge ($Z_P$), 
target charge ($Z_T$) and $E$ dependencies. As a result theoretical estimates of the asymptotic ($E \rightarrow \infty$)
energy dependence of the total cross sections are not in agreement, with
estimates of \cite{bert88,baltz91}: 
$\sigma_{\rm CPP} \sim \ln (E) $, and more recently \cite{eich95}: $\sigma_{\rm
CPP} \sim E^2 $. The former is based on the positron-electron pair being created
around the same ion, the latter assuming that the pair is divided between the two ions.
Of course both pathways will interfere and  
contribute to the process,
thus pointing to the necessity of a two-center treatment for
the positron and electron. It has been shown \cite{deco88b} that the 
two-center description is essential in obtaining the correct positron emission 
spectrum and accurate total cross section for CPP. 
Leading-order perturbation theory (the first Born approximation) does give 
reasonably good estimates for the cross section in the 
high-energy region ($E \sim 150 \ {\rm GeV/u}$) 
\cite{krau98} for collisions of heavy ions, and has been a reliable model
for  fast collisions of light ions with low $Z$ targets  
in the process of antihydrogen formation 
involving CPP by antiprotons \cite{baur96,bert98}. 

Experimental results for  highly relativistic heavy ions on a variety of
targets \cite{belk98} support the simple scaling law derived from the virtual-photon
method (Born approximation) which included multiple scattering from the 
projectile ion \cite{bert88} alone:
$\sigma_{\rm CPP} \sim Z_T^2 $, for a given energy. At lower energies 
this is not the case \cite{belk94a,belk94b,belk97}, the $Z_T$ dependence is more complex, showing
an enhancement in excess of the $Z_T^2$-scaling.
In this paper we propose
a refinement of the Born approximation to take into account higher-order scattering
processes. In particular we tackle the question of the two-center nature of the
continuum positron and the polarization of the captured electron. We find both these
effects are vital and lead to theoretical results which are in accord with experiment.
We discuss the physical explanation for scaled cross section enhancement 
and  provide numerical estimates which agree
very well with experiment in qualitative and quantitative terms. 

Through crossing symmetries the leading-order matrix element for the pair production process, in which the electron is captured by the projectile $P$,
\begin{equation}
 P + T \rightarrow (P,e^{-}) + T + e^{+}
\label{cpp}
\end{equation}
is the same as that for the related reaction,
$\ \ e^{-} + P + T \rightarrow (P,e^{-}) + T \ \ $,
which is mathematically equivalent to the time-reversed ionization process
\begin{equation}
(P,e^{-}) + T \rightarrow e^{-} + P + T .
\label{ion}
\end{equation}
In each crossing symmetry the equivalence relies on the electron-positron interaction being much weaker
than their interactions with the highly-charged ions; a reasonable assumption.
Let $\mbox{\boldmath $ r$}_P, t$ and $\mbox{\boldmath $ r$}'_T, t'$ be the space and time
coordinates of the electron in the projectile and target frames, respectively.
The nuclei follow straight-line paths with relative
velocity $\mbox{\boldmath $v$}$.
The Hamiltonian, in the projectile frame of reference and in 
atomic units, is given by:
\begin{equation}
H = -ic{\BMalpha}.{\BMnabla}_{\mbox{\boldmath $r$}_P}\ + \beta c^{2} + 
V_{P}(\mbox{\boldmath $r$}_P) + S^{2}V'_T(\mbox{\boldmath $ r$}'_T) 
\end{equation}
where $\BMalpha$ and $\beta$ are Dirac matrices and $S$ is the operator
which transforms the wavefunction from the projectile frame to the target
frame, namely
\begin{equation}
 S = ({\textstyle{1 \over 2}}+ {\textstyle{1 \over 2}} \gamma)^{\frac{1}{2}}
({\bf 1}-x{\BMalpha}.\widehat{\mbox{\boldmath $ v$}} ),
\end{equation}
with $x=v\gamma c^{-1}(\gamma+1)^{-1}, \gamma=(1-v^2/c^2)^{-{1}/{2}}$, and {\bf 1}
represents the unit matrix.
For a given impact parameter $\mbox{\boldmath $ b$}$, 
the transition amplitude can be written in the form \cite{croth83}
\begin{equation}
 A(\mbox{\boldmath $ b$}) = -i\int_{-\infty}^{\infty}dt \int d \mbox{\boldmath $r$}_P
\ \ \chi^{\dag}_{f} (H-i\partial_t) \chi_{i},
\label{amp} 
\end{equation}
where $\chi_{i}$ and $\chi_f$ are the initial and final states.

The  undistorted  bound-state is approximated by a semirelativistic 
($Z_T \ll c$) wavefunction :
\begin{equation}
 \Phi_i=\Phi_{0i}+\Phi_{1i},
\end{equation}
where
\begin{equation}
\Phi_{0i}=Z_{P}^{\frac{3}{2}}\pi^{-\frac{1}{2}}e^{-Z_{P}r_P-ic^2t-iE_{si}t}{\BMomega}_i
\end{equation}
and
\begin{equation}
\Phi_{1i}=(2ic)^{-1}{\BMalpha}.{\BMnabla}_{{\bf r}_{P}} \Phi_{0i},
\end{equation}
with $E_{si}$ the non-relativistic  eigenenergy, and the electron 
spin along the beam axis defined as `up' by: 
${\BMomega}_{i}^T=( 1 \; 0 \; 0 \; 0)$ and `down' by ${\BMomega}_{i}^T=( 0 \; 1 \; 0 \; 0)$

The continuum function is given by:
\begin{equation}
\Phi_f=\Phi_{0f}+\Phi_{1f},
\end{equation}
where
\begin{eqnarray}
\Phi_{0f}&=&(2\pi)^{-\frac{3}{2}}N^*(\omega_P)
_1F_1(-i\omega_P;1;-i\gamma_e(v_e r_P+ \mbox{\boldmath $ v$}_e \cdot \mbox{\boldmath $ r$}_P)) 
\nonumber \\
 &&\times \ e^{-i\gamma_e c^2 t+i\gamma_e \mbox{\boldmath $ v$}_e \cdot \mbox{\boldmath $ r$}_P}
S_{{\bf v}_e}^{-1}{\BMomega}_f. 
\end{eqnarray}
The spinor correction term is given by
\begin{eqnarray}
\Phi_{1f}&=& (2\pi)^{-\frac{3}{2}}(2i\gamma_ec)^{-1}N^*(\omega_P) \ \times 
\nonumber \\
&&{\BMalpha}.{\BMnabla}_{r_P1}
F_1(-i\omega_P;1;-i\gamma_e(v_er_P+\mbox{\boldmath $ v$}_e.{\bf r}_P)) 
\nonumber \\
&& \qquad \times \ e^{-i\gamma_ec^2t+i\gamma_e{\bf v}_e.{\bf r}_P}
S_{\mbox{\boldmath $v$}_e}^{-1}{\BMomega}_f . 
\end{eqnarray}
with ${\omega}_P=Z_P/v_e$, where $\mbox{\boldmath $v$}_e$ is the electron velocity.
$N(\zeta)=\exp\left({\pi\zeta}/{2}\right)\Gamma(1-i\zeta)$ and,
\begin{equation}
 S_{\mbox{\boldmath $v$}_e} = ({\textstyle{1 \over 2}}+ 
{\textstyle{1 \over 2}} \gamma_e)^{\frac{1}{2}}
({\bf 1}-x_e{\BMalpha}.\widehat{\mbox{\boldmath $ v$}}_e ),
\end{equation}
where $x_e=v_e\gamma_e c^{-1}(\gamma_e+1)^{-1}$, and $ \gamma_e=(1-v_e^2/c^2)^{-{1}/{2}}$.
These functions are appropriate when $ Z_{P,T} \ll c$.

The initial distortion factor $\mbox{\boldmath $ L$}'_i$ is a matrix given by:
\begin{equation}
\mbox{\boldmath $ L$}'_i=\mbox{\boldmath $ L$}'_{0i}+ \mbox{\boldmath $ L$}'_{1i},
\end{equation}
where
\begin{equation}
\mbox{\boldmath $ L$}'_{0i}= 
\exp( -i\nu_T \ln [\gamma v r_T'+\gamma \mbox{\boldmath $ v$} \cdot \mbox{\boldmath $ r$}'_T ] ){\bf 1},
\end{equation}
and
\begin{equation}
\mbox{\boldmath $L$}'_{1i}=S^{-1} (2i\gamma c)^{-1}{\BMalpha}.{\BMnabla}_{r'_T}
\mbox{\boldmath $L$}_{0i} S,
\label{l1i}
\end{equation}
with $\nu_{T}=Z_T/v$.

The final state distortion is given by \cite{deco88b}:
\begin{equation}
\mbox{\boldmath $ L$}'_f=\mbox{\boldmath $ L$}'_{0f}+ \mbox{\boldmath $ L$}'_{1f},
\end{equation}
where
\begin{equation}
\mbox{\boldmath $ L$}'_{0f}=N^*(\omega'_T)
_1F_1(-i\omega'_T;1;-i\gamma_e'(v_e'r_T'+\mbox{\boldmath $ v$}'_e \cdot
 \mbox{\boldmath $ r$}'_T)){\bf 1}
\end{equation}
and
\begin{equation}
\mbox{\boldmath $L$}_{1f}'= S^{-1}(2i\gamma_e'c)^{-1}{\BMalpha}.{\BMnabla}_{r_T'}
 \mbox{\boldmath $L$}_{0f}' S.
\end{equation}

Retaining  terms of first order in $Z/c$, we have relativistic continuum distorted wave eikonal initial state (RCDWEIS)
wavefunctions \cite{croth83,deco88a}:
\begin{equation}
\chi_i=\mbox{\boldmath $ L$}_{0i}'\Phi_{0i}+\mbox{\boldmath $L$}_{1i}'\Phi_{0i}+
\mbox{\boldmath $ L$}_{0i}' \Phi_{1i},
\label{rcdwi}
\end{equation}
\begin{equation}
\chi_f= \mbox{\boldmath $ L$}_{0f}'\Phi_{0f}+ \mbox{\boldmath $L$}_{1f}'\Phi_{0f}
+\mbox{\boldmath $L$}_{0f}'\Phi_{1f}.
\label{rcdwf}
\end{equation}


\begin{figure}
\centering\epsfig{file=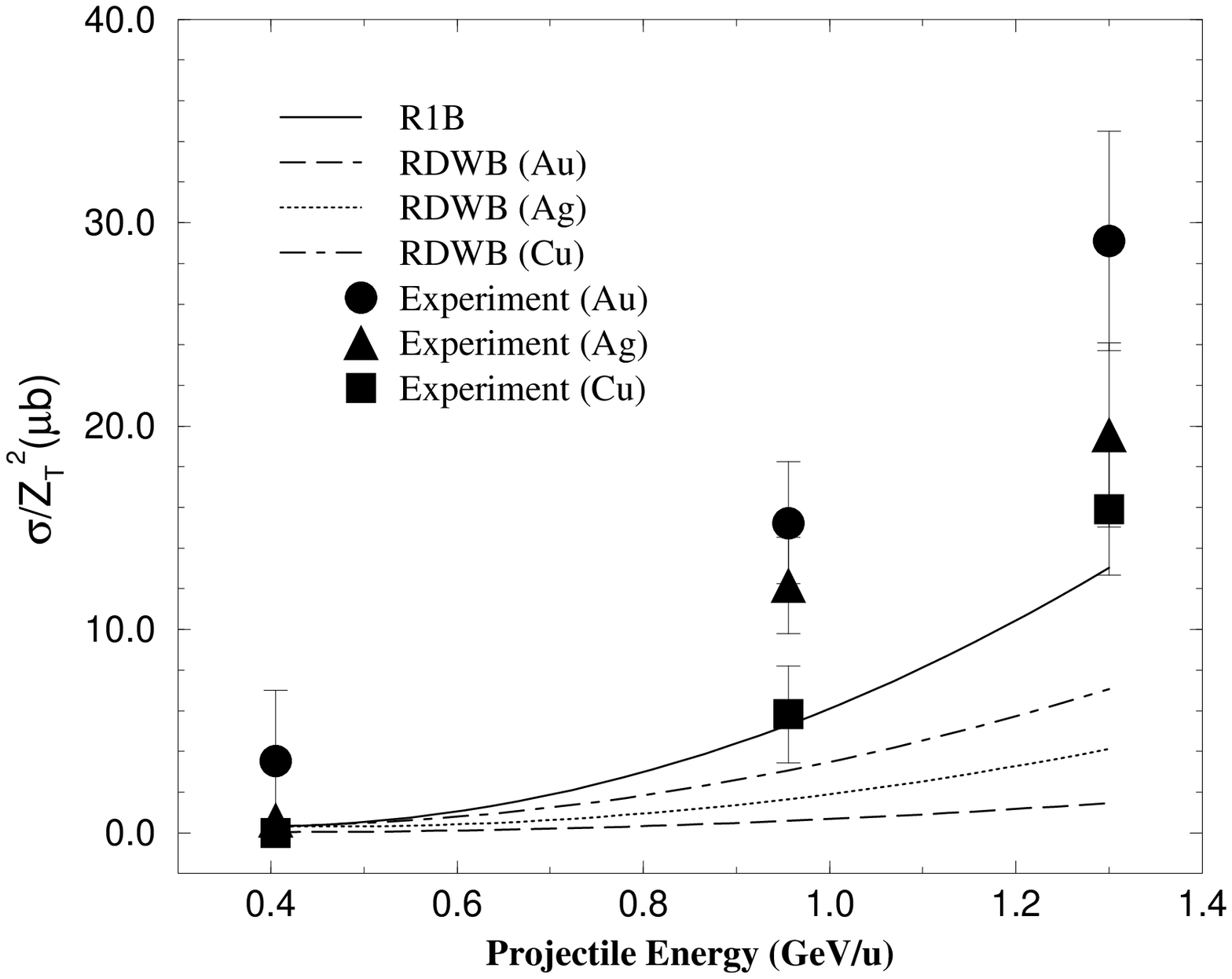, clip=, width=7.25cm, 
 angle=0}
\caption{Scaled cross sections, $\sigma_{\rm CPP}/Z_T^2$ in
microbarns, for pair-production with electron capture by 
fully stripped Lanthanum
ions (${\rm La}^{57+}$)
striking thin foils of Copper ($Z_T=29$), Silver ($Z_T=47$) and Gold ($Z_T=79$).
Comparison with RDWB theory for capture to the $1s$-state.  
 }
\label{figure1}
\end{figure}

\begin{figure}
\centering\epsfig{file=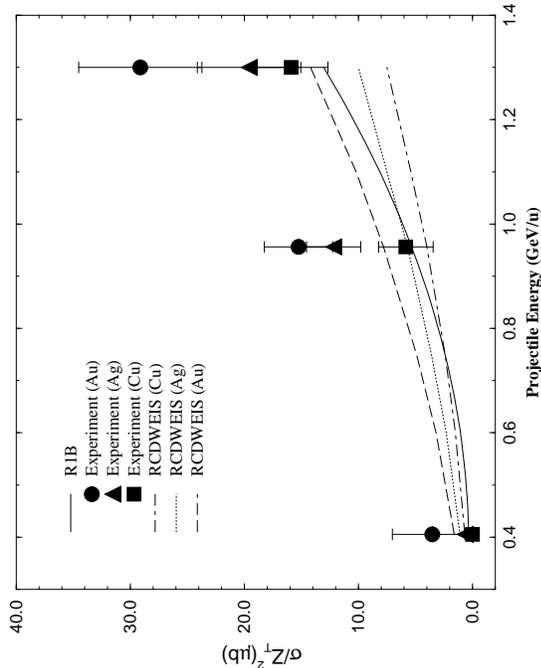, clip=, width=7.25cm, 
 angle=0}
\caption{Scaled cross sections, $\sigma_{\rm CPP}/Z_T^2$ in
microbarns, for pair-production with electron capture by
fully stripped Lanthanum ions (${\rm La}^{57+}$)
striking thin foils of Copper ($Z_T=29$), Silver ($Z_T=47$) and Gold ($Z_T=79$).
Comparison with RCDWEIS theory for capture to the $1s$-state.  
 }
\label{figure2}
\end{figure}

We compare our results for the Relativistic First Born (R1B), projectile centered, approximation,
in which the initial and final state
distortions are neglected, and the Relativistic Distorted Wave Born (RDWB)
approximation \cite{deco88b}, where the two-center positron wavefunction
is used but 
the initial state distortion is omitted. The Born approximation, which assumes that 
the positron is in the continuum of only one of the ions, is ambiguous. The 
Born approximation of Bertulani and Baur \cite{bert88} assumes the positron to be a projectile-centered
wavefunction, while the Born approximation of Eichler \cite{eich95} takes the 
positron to be centered around the target. These two models can be viewed as approximations to the wavefunction (\ref{rcdwf}) 
in which $\omega'_T=0$ and $\omega_P=0$, respectively. By 
retaining both scattering center contributions the interference effects are 
taken into account.  In comparing RDWB and R1B, it
is known that {\it these} two-center 
interference effects reduce the cross section for CPP in the relativistic 
domain \cite{deco88b}. This suppression of CPP is the converse of the
two-center enhancement (capture to the continuum) 
that arises in ion-atom ionization \cite{croth83} 
and is analogous to 
the effect of the  Fermi function for $\beta^{\pm}$ decay \cite{enge78}.

The triply differential cross section, with respect to the electron momentum
($\mbox{\boldmath $p$}_e$) is defined as: 
\begin{equation}
 \sigma( \mbox{\boldmath $ p$}_e)=(d \sigma_{\rm CPP}/d {\mbox{\boldmath $ p$}}_e)
 = \int d \mbox{\boldmath $b$}
 |A( \mbox{\boldmath $ b$})|^2 .
\end{equation}

Using the Fourier transform method \cite{croth83} we define
\begin{equation}
T( \mbox{\boldmath $ \eta$}) = \gamma v \int d \mbox{\boldmath $ b$}\  
\exp(-i \mbox{\boldmath $\eta$} \cdot \mbox{\boldmath $ b$})  A( \mbox{\boldmath $b$}),
\end{equation}
where $T({\BMeta})$ is a product of single-center integrals.
The total cross section is obtained from the integral over the ejectile 
momentum (or velocity) and takes the form: 
\begin{equation}
 \sigma_{\rm CPP} = \sum_{\rm spins} \frac{1}{2\pi (\gamma v)^2}\int^c_0 dv_e \gamma_e^5v_e^2\int_0^{\pi}d\theta \sin \theta \int d\mbox{\boldmath $\eta$} 
 |T({\BMeta})|^2 .
\end{equation}
where we sum over all the spin states of the electron and positron pair.

In order to compute CPP cross sections (\ref{cpp}) we note that 
a positron with energy $\epsilon_+$ and momentum 
${\bf p}_+$ traveling forward in time in the final state is equivalent to an electron with energy $-\epsilon_+$ and momentum $-{\bf p}_+$ 
in the initial state. Thus we must take:
\begin{eqnarray}
v_e \rightarrow -v_+   & \ \ \ \ 
v_{e}' \rightarrow -v_{+}' \nonumber \\
\epsilon_f \rightarrow -\epsilon_+  & \ \ \ \ 
\epsilon_{f}' \rightarrow -\epsilon_{+}' .
\end{eqnarray}

The experiments of Belkacem {\it et al.} \cite{belk94a,belk94b,belk97} were for fully stripped Lanthanum
ions (${\rm La}^{57+}$)
striking thin foils of Copper ($Z_T=29$), Silver ($Z_T=47$) and Gold ($Z_T=79$).
The collision energies were $E=0.405, 0.956 \ {\rm and}\ 1.300\ $ GeV/u. 
The two graphs presented  compare the scaled total cross sections ($\sigma_{\rm CPP}/Z_T^2$) 
given by theory and experiment. Consider figure \ref{figure1} which compares 
R1B and RDWB with the measurements. Of course, the scaled R1B curve is independent of $Z_T$, 
and it clearly shows the increase in importance of CPP with increasing collision energy.
Considering the RDWB model however,
we see a progressive  reduction in the scaled cross section as $Z_T$ increases.
This is in agreement with the findings of Deco and Rivarola \cite{deco88b}, 
who reported  a decrease in the size of the singly differential cross sections by an order of magnitude. 
Their model is essentially our RDWB approximation. While this model shows $Z_T$ dependence for the 
scaled cross section, the trends and absolute values are incorrect. It predicts a suppression of the scaled
cross section rather than an enhancement as $Z_T$ increases. Thus the
RDWB theory data for Gold gives the lowest scaled cross-section
while experiment shows that it should be the highest.
The same incorrect trend was obtained by the target-centered
Born approximation \cite{eich95}.

In contrast (figure \ref{figure2}) the equivalent results for RCDWEIS
 show the observed enhancement with
increasing  $Z_T$. However the theoretical data lie below the 
experiment for the more energetic collisions. In comparing
with experiment we have only presented simulations for the
dominant channel; capture to the $1s$ ground state. At very high
energies  capture to excited states  
is thought to contribute $\sim 30\%$ to the total capture cross section
\cite{krau98,baltz96}.  This would partly explain the differences between 
our theory results and the experimental data. 
These results are a great improvement on the R1B approximation and are consistent with 
the experimental data.

Other experimental results are available for the impact of faster and more
highly charged beams: 10.8  GeV/u 
${\rm Au}^{79+}$ \cite{belk98} and $0.956$ GeV/u ${\rm U}^{92+}$ \cite{belk93} 
for the same targets.
The Gold beam results (table \ref{table1})
 indicate that the $Z_T^2$ dependence is established at the
higher energies, as predicted by the simple projectile-centered
Born approximation \cite{beck87a}. Even at this higher 
energy our theoretical results (table \ref{table1}) 
show an enhancement in excess  of $ Z_T^2$. The experiment is
in much better accord with the flat scaled cross section data 
given by the Born approximation \cite{beck87a}. 

Finally we note that for very large charges, for example ${\rm U}^{92+}$,
the semirelativistic approximations for the wavefunctions are not valid.
Qualitatively this can lead to
underestimation of the cross sections for high charges when using
approximate wavefunctions \cite{beck87b,eich95}.
In such cases, Coulomb-Dirac wavefunctions  are
required in order to model the process \cite{beck87b}. 

In summary, we have proposed and tested a new distorted wave model
which incorporates and reduces to 
approximations used previously to describe CPP. We confirm that, as has previously been shown \cite{deco88b}, 
the inclusion of distortions from both ions on
the positron continuum state leads to a reduction in the cross sections. However including distortion of the bound electron leads to an increase in the
total cross sections and a more accurate fit to the experimental data for
fully-stripped relativistic Lanthanum ions.
This improvement in results when including distortions demonstrates once more the necessity of a two-center treatment for an accurate theoretical
 description of this reaction.   However our 
cross section predictions for faster 
and more highly charged Gold ions  do not accord with the experimental
data which show a $Z_T^2$ dependence.	While the refinements
introduced in our model are significant theoretical improvements,
 clearly there still exist 
several unresolved important differences between theory and experiment.

\acknowledgements{R.J.S. Lee  and J.V. Mullan acknowledge financial
support from the Department of Education for Northern Ireland 
through the Distinction Award scheme. }

\begin{table}
\caption{Total cross sections, $\sigma_{\rm CPP}$ in barns, for electron capture from pair production for
10.8 GeV/nucleon Au$^{79+}$ impact on Gold, Silver and Copper foils.}
\begin{tabular}{l l l l}
Z$_T$ & Experiment \cite{belk98} & CDWEIS theory & Becker {\it et al} 
\cite{beck87a} \\ \hline
79 &  8.8 $\pm$ 1.5    & 15.85 &  10.1  \\
47 &  4.4 $\pm$ 0.73   & 3.44  &   3.6  \\
29 &  1.77 $\pm$ 0.31  & 0.74  &   1.36 \\
\end{tabular}
\label{table1}
\end{table}

\end{document}